\documentstyle[12pt]{article}
\begin{document}
\title{First class constraints in Regge calculus}
\author{V. Khatsymovsky \\
 {\em Institute of Theoretical Physics} \\
 {\em Box 803} \\
 {\em S-751 08 Uppsala, Sweden\thanks{Permanent adress (after 15
November
1993): Budker Institute of Nuclear Physics, Novosibirsk 630090,
Russia}} \\
 {\em E-mail address: khatsym@rhea.teorfys.uu.se\thanks{Permanent
E-mail
address (after 15 November): khatsym@inp.nsk.su}}}
\date{\setlength{\unitlength}{\baselineskip}
\begin{picture}(0,0)(0,0)
\put(9,13){\makebox(0,0){UUITP-25/93}}
\end{picture}
}
\maketitle
\begin{abstract}
Considered are I class constraints in the tetrad-connection
formulation of
Regge calculus. One of these is well-known Gauss law which generates
rotations
in the local frames associated with tetrahedrons in the continuous
time 3D
section. Another two types of these are new, satisfied by definition
of Regge
manifold and having no I class analogs in the continuum general
relativity.
Constraints of the first type express vanishing of the dual squares
of
antisymmetric tensors of the triangles in the 3D section thus
ensuring each
such tensor being a bivector. Constraints of the second type are
trigonometric
relations between areas of triangles of 3D section caused by that the
set of
areas is redundant as compared to the set of linklengts.
\end{abstract}
\newpage

When passing from continuum to discrete theory continuous symmetries
are
usually lost. The more surprising is that Regge calculus (suggested
by Regge
\cite{Regge}; see recent review \cite{Will-rev}) or, more exactly,
it's
continuous time tetrad-connection formulation \cite{Kha} possesses
some
symmetries having no analogs in the continuum general relativity (GR)
in the
Palatini form (see, e.g.,  \cite{Rom}). These symmetries are, in
fact, remnants
of the more rich symmetry w.r.t. rotations in the 4-tetrahedrons of
the
original completely discrete Regge manifold before passing to the
continuous
time limit.

Let us now describe Regge Lagrangian of \cite{Kha} in the more
detailed form
given in \cite{Kha1} where also detailed consideration of the
canonical form of
Regge calculus is given. To make one of coordinates, just called time
$ t$,
continuous we should tend dimensions of simplices in definite
direction to
zero. For that we consider the set of all the Regge vertices being
groupped in
3D manifolds $ t=const$ (leaves) distinguished by the distances
tending to zero
in the continuous time limit. Denote by $ i,~k,~l,...$ vertices of
the given
leaf at $ t$; next leaf at $ t+dt$ consists of vertices $
i^+,~k^+,~l^+,...$.
Denote $n$-simplex by unordered sequence of it's vertices $
(A_1...A_{n+1})$
where round brackets mean unordering. Here $A=i$ or $i^+$, that is,
we define
triangulation so that only simplices appear which are completely
contained
either in the one leaf (call these leaf simplices) or between the two
neighboring ones. The latter we call infinitesimal if their
$n$-volume is
$O(dt)$ or diagonal if it is finite, $O(1)$. By definition, first,
mutual
correspondence $ i\leftrightarrow i^+$ exists and links $(ii^+)$ are
put to be
infinitesimal and, second, links $(ik)$ and $(i^+k^+)$ either exist
or do not
simultaneously.

Thus, Regge manifold between each two neighboring leaves became the
set of
infinitesimal 4D prisms with tetrahedrons $(iklm)$ as bases.
Originally in the
full discrete Regge manifold local frames were attributed to the
4-tetrahedrons; now we refer them to 4-prisms or rather to
3-tetrahedrons at
the given moment $t$. Also originally we had connections $\Omega$
(SO(3,1)
matrices) on 3-tetrahedrons $(ABCD)$; now those on the leaf and
diagonal
tetrahedrons are infinitesimal, e.g.

\begin{equation}\label{f}
\Omega_{(iklm)}={\bf 1}+f_{(iklm)}dt,~~~\bar{f}=-f,
\end{equation}
and those on infinitesimal 3-tetrahedrons are finite:

\begin{equation}\label{}
\Omega_{(ii^+kl)}\stackrel{\rm def}{=}\Omega_{i(kl)}.
\end{equation}
Frame indices are $a,~b,~c,...=0,~1,~2,~3$, signature is (-,+,+,+).
Of these
matrices we form curvatures which are path ordered products of
$\Omega$'s
around triangles $(ABC)$ and substitute them into 4D Regge action,

\begin{equation}\label{S}
S=\sum_{(ABC)}|v_{(ABC)}\!|\arcsin\frac{v_{(ABC)}}{|v_{(ABC)}\!|}\circ

R_{(ABC)},~~~(|v|\stackrel{\rm def}{=}(v\circ v)^{1/2},~~~P\circ
Q\stackrel{\rm
def}{=}\frac{1}{2}P^{ab}Q_{ab}).
\end{equation}
Here $v_{ab}$ is antisymmetric area tensor,

\begin{equation}\label{v-l-l}
v_{ab}=\epsilon_{abcd}l^c_1l^d_2,
\end{equation}
(bivector) for a triangle formed by vectors $l^a_1,~l^a_2$; it will
be denoted
as $\pi_{(ikl)}$ for the leaf triangle $(ikl)$ (and thus for a
diagonal one
close to that up to $O(dt)$) and $n_{ik}dt$ for infinitesimal
triangle
$(ii^+k)$ or $(ii^+k^+)$. Substitute this into (\ref{S}) and pass to
integration over $dt$. Contributions of the leaf and diagonal
triangles give
the following terms in the Lagrangian: kinetic

\begin{equation}\label{kin}
L_{\dot{\Omega}}=\sum_{(ikl)}\pi_{(ikl)}\circ(\bar{\Omega}_{(ikl)}\dot
{\Omega}_{(ikl)}),
\end{equation}
Gaussian,

\begin{equation}\label{Gauss}
L_h=\sum_{(iklm)}h_{(iklm)}\circ\sum_{{\rm cycle\, perm}\, iklm}
\varepsilon_{(ikl)m}\Omega^{\delta_{(ikl)m}}_{(ikl)}\pi_{(ikl)}
\Omega^{-\delta_{(ikl)m}}_{(ikl)},~~~
(\delta \stackrel{\rm def}{=}\frac{1+\varepsilon}{2}),\nonumber\\
\end{equation}
and new one, $L_{\phi}$. Definition of $\Omega_{(ikl)}$ and
$L_{\phi}$ is given
below. Here $h_{(iklm)}$ is the sum of matrices of the type of
$f_{(iklm)}$ for
the four successive in time leaf and diagonal tetrahedrons in the
4-prism
$(iklm)$, the true analog of time component of connection
$\omega^{ab}_0$ in
the continuum GR, and $ \varepsilon_{(ikl)m}=\pm 1$ is sign function
which
specifies for each bivector $ \pi$ choice of one of two tetrahedrons
in the
frame of which it is defined. Contribution of infinitesimal triangles
gives
induced 3D curvature term

\begin{eqnarray}\label{L-n}
L_n & = & \sum_{ik}|n_{ik}\!|\arcsin\frac{n_{ik}}{|n_{ik}\!|}\circ
R_{ik}\\
&&(R_{ik}=\Omega^{\varepsilon_{ikl_n}}_{i(kl_n\!)}\ldots
\Omega^{\varepsilon_{ikl_1}}_{i(kl_1\!)},~~~ \varepsilon_{ikl_j}\!=
-\varepsilon_{(ikl_j)l_{j-1}}\!=\varepsilon_{(ikl_j)l_{j+1}},~~~|n|\st
ackrel{\rm def}{=}(n\circ n)^{1/2}).\nonumber
\end{eqnarray}

Consider now $L_\phi$. The $L_{\phi}dt$ term in the action arises as
a result
of cancellations between finite contributions to action of the leaf
and
diagonal triangles of 3-prisms, see Fig.1.

\begin{eqnarray}
\label{prism}
\begin{picture}(60,130)(50,0)
\put(-100,90){\rm Fig.1. Infinitesimal 3-prism}
\put (100,20){\line(0,1){80}}
\put (100,20){\line(1,1){120}}
\put (100,20){\line(2,1){160}}
\put (100,20){\line(3,1){120}}
\put (100,20){\line(1,0){160}}
\put (100,100){\line(3,1){120}}
\put (100,100){\line(1,0){160}}
\put (220,60){\line(0,1){80}}
\put (220,60){\line(1,-1){40}}
\put (220,140){\line(1,-1){40}}
\put (220,140){\line(1,-3){40}}
\put (260,20){\line(0,1){80}}
\put (260,95){$~l^{+}$}
\put (220,140){$~k^{+}$}
\put (87,95){$i^{+}$}
\put (92,15){$i$}
\put (220,60){$~k$}
\put (260,15){$~l$}
\end{picture}\nonumber
\end{eqnarray}
Such the cancellation results from specific form of $\Omega_{i(kl)}$
following
from eqs. of motion. Indeed, $O(1)$ part of curvature on, say,
triangle
$(ik^+l)$ takes the form

\begin{equation}
R_{(ikk^+l)}=\bar{\Omega}_{k(li)}\Omega_{l(ik)}.
\end{equation}
Upon variation of connection on the infinitesimal tetrahedron,

\begin{equation}
\delta\Omega_{l(ik)}=w_{l(ik)}\Omega_{l(ik)}dt,~~~\bar{w}=-w,
\end{equation}
finite variation of $L$ linear in $w$ just arises from $(ik^+l)$ and
$(ik^+l^+)$ contributions into action. Solution to this constraint
and those
with permuted $i,~k,~l$ reads

\begin{equation}
\Omega_{i(kl)}=\Omega_{(ikl)}\exp(\phi_{i(kl)}\pi_{(ikl)}+
\,^{*}\!\phi_{i(kl)}\,^{*}\!\pi_{(ikl)}),~~~^{*}\!P_{ab}\stackrel{\rm
def}{=}{1
\over 2}\epsilon_{abcd}P^{cd},
\end{equation}
$\phi,~^{*}\!\phi$ being new independent variables. Appearing here
$\Omega_{(ikl)}$ just stands in the $L_{\dot{\Omega}},~L_h$ above. So
we get
contribution of the triangle $(ik^+l)$ proportional to
$\phi_{l(ik)}-\phi_{k(li)}$ which vanishes being summed over
permutations of
$i,~k,~l$ (that is, over $(ikl)$, $(ik^+l)$, $(ik^+l^+)$, see Fig.1).
Important
is that these $O(1)$ contributions enter action multiplied by up to
$O(dt)$ the
same area factor $|\pi_{(ikl)}|^2$; on the $O(dt)$ level one should
take into
account corrections to it of the type $\pi\circ\delta\pi=\pi\circ
ndt$
(symbolically, $ndt$ is total lateral area tensor of infinitesimal
3-tetrahedrons between these triangles). This gives new term

\begin{equation}
L_{\phi}=-\sum_{(ikl)}\pi_{(ikl)m}\circ\sum_{{\rm perm}\, ikl}
\varepsilon_{ikl}\phi_{i(kl)}n_{ik(lm)}
\end{equation}
Here more detailed notations $ \pi_{(ikl)m}$ and $ n_{ik(lm)}$ show
that given
quantities are defined in the frame of the tetrahedron $ (iklm)$.
This implies
knowledge of $\pi_{(ikl)},~n_{ik}$ more than in one frame. Therefore
we define
$ \pi,~n$ in the frames of all the 4-tetrahedrons containing
corresponding
triangles simultaneously imposing a set of constraints.

These kinematical (i.e. imposed by definition of Regge manifold)
constraints
are structurally rather simple (bilinears in $\pi$, $n$) and can be
conventionally divided into two groups: dual and scalar ones. Dual
constraints
express either vanishing dual product,
$$P*Q\stackrel{\rm def}{=}\frac{1}{4}\epsilon_{abcd}P^{ab}Q^{cd},$$
of two bivectors of triangles if these triangles have common edge or
equality
of these products up to sign inside the 4-tetrahedron otherwise (in
this case
it equals, up to sign, to the tetrahedron's 4-volume). Dual
constraints ensure
that there are 4 4-vectors which span the given tetrahedron in terms
of which
it's 2-face bivectors are expressed like (\ref{v-l-l}), analogously
as
bivectors in the continuum GR in terms of tetrad  are; see, e.g.,
\cite{Rom}.

Scalar constraints are specific for Regge calculus and serve to
guarantee
unambiguity of the linklength calculated in the different
4-tetrahedrons. It is
sufficient to require for that unambiguity of scalar products $
v_{(ABC)}\circ
v_{(ABD)}$ for 2-faces of each 3-tetrahedron $(ABCD)$. Indeed, the
number 6 of
independent scalar products for it coincides with the number of
3-tetrahedron
links.

Consider now possible symmetries of the Lagrangian. Since variation
of any
linklength generally means change of geometry, it is natural to
assume that
symmetries should arise only as consequence of original SO(3,1)
rotational
symmetry in the 4-tetrahedrons. Now we have rotations in the 4-prisms
and also
some remnants of original symmetry w.r.t. the rotations in {\it
separate}
4-tetrahedrons; the latter should involve some transformations of $
\phi,~^{*}\!\phi$.

Now if we define Poisson brackets $\{\cdot,\cdot\}$ so that for any
function
$f(\pi,\Omega)$ we would have

\begin{equation}
\frac{df}{dt}=\{f,H\},~~~L\stackrel{\rm def}{=}L_{\dot{\Omega}}-H
\end{equation}
then these should be

\begin{equation}
\label{{}}
\{f,H\}=\sum_{\rm triangles}^{}{\pi\circ
[H_{\pi},f_{\pi}]+H_{\pi}\circ
\bar{\Omega}
f_{\Omega}-f_{\pi}\circ \bar{\Omega}H_{\Omega}}.
\end{equation}
(indices $\pi,~\Omega$ mean differentiation). It is easy to check
that via
these brackets $L_h$ acts as generator of local rotations in the
4-prisms, $h$
being infinitesimal parameter of these rotations, and forms usual
algebra\footnote{Of course, the same can be obtained in the more
rigorous way
introducing, as in \cite{Wael}, the variables $
P_{ab}=\Omega_a^{~c}\pi_{cb}$
and $ \Omega^{ab}$, canonically conjugated in usual sense. There will
be II
class constraints to which $ P,~\Omega$ are subject, and $
\{\cdot,\cdot\}$
turn out to be Dirac brackets w.r.t. this system of II class
constraints.}.

Besides that, there is invariance under symmetrical in $ i,~k,~l$
variation of
$^*\!\phi$:

\begin{equation}\label{d*phi}
\delta\Omega_{(ikl)}=\Omega_{(ikl)}\,^*\!\pi_{(ikl)}\lambda_{(ikl)},~~
{}~\delta\,^*\!\phi_{i(kl)}=-\lambda_{(ikl)},~~~{\rm
cycle~perm}~i,~k,~l,
\end{equation}
$\lambda_{(ikl)}$ being infinitesimal parameter. Indeed, upon such
variation
Lagrangian acquires, up to the full derivative, additional term

\begin{equation}
\delta L={1 \over 2}\dot{\lambda}_{(ikl)}\pi_{(ikl)}*\pi_{(ikl)}.
\end{equation}
This is just dual constraint ensuring vanishing the dual square of
bivector; at
the same time it generates (\ref{d*phi}).

As for the analogous variation of $ \phi$,

\begin{equation}
\delta\Omega_{(ikl)}=\Omega_{(ikl)}\pi_{(ikl)}\lambda_{(ikl)},~~~\delt
a\phi_{i(kl)}=-\lambda_{(ikl)},~~~{\rm cycle~perm}~i,~k,~l,
\end{equation}
it leads to appearance in $ L$ of the term with derivative of the
squared area,
$ \pi\circ\dot{\pi}$. The idea is to convert this area derivative
into the
derivative of a constraint and, adding full derivative, transfer
differentiation from $ f$ to $ \lambda$. For that $ f$ should be
purely
function of areas, and then we put

\begin{equation}
\lambda_{(ikl)}=\lambda\partial f/\partial(|\pi_{(ikl)}|^2).
\end{equation}
The result reads

\begin{equation}
\delta L={1 \over 2}\dot{\lambda}f+\lambda\sum_{(ikl)}^{}{{\partial f
\over
\partial (|\pi_{(ikl)}|^2}\left[ \pi_{(ikl)}\circ\sum_{{\rm perm}\,
ikl}^{}{\varepsilon_{ikl}}n_{ik} \right]}.
\end{equation}
Surrounded by square brackets is $ \pi\circ\delta\pi=\delta|\pi|^2/2$
expressed
in terms of lateral area tensor. Therefore second term in $ \delta L$
is
proportional simply to variation of the constraint $ f$ when passing
from $ t$
to $ t+dt$ leaf, i.e. again to the kinematical constraint. Thus $
\delta L$ is
combination of constraints. The only question remains whether such
function $
f$ does exist. The answer is confirmative, because the number of
areas is
larger than the number of lengths in 3D leaf. This is easily to see
since each
triangle has three edges but each edge is shared by no less than
three
triangles. Therefore areas should satisfy a number of additional
conditions in
order to be functions of linklengths. As we have shown just now,
these
conditions are I class constraints.

\bigskip
I am grateful to prof. A. Niemi, S. Yngve and personnel of Institute
of
Theoretical Physics at Uppsala University for warm hospitality and
support
during the work on this paper.


\begin{thebibliography}{99}
\bibitem{Regge}
 T.Regge, {\it Nuovo Cim.~}{\bf 19}~(1961)~558
\bibitem{Will-rev}
 R.M.Williams and P.A.Tuckey, {\it Class.Quantum Grav.~}{\bf
9}~(1992)~1409
\bibitem{Kha}
 V.Khatsymovsky, {\it Class.Quantum Grav.~}{\bf 6}~(1989)~L249;~{\bf
8}~(1991)~1205
\bibitem{Rom}
 J.D.Romano, {\it Gen.Rel.Grav.~}{\bf 25}~(1993)~759
\bibitem{Kha1}
 V.Khatsymovsky, {\it Regge calculus in the canonical form}
Novosibirsk
preprint BINP 93-42 (1993)
\bibitem{Wael}
 H.Waelbroeck, {\it Class.Quantum Grav.~}{\bf 7}~(1990)~751
\end{thebibliography}
\end{document}